\documentclass[12pt]{iopart}

\newcommand{\vect}[1]{{\bf #1}}
\renewcommand{\d}{{\rm d}}
\renewcommand{\i}{{\rm i}}
\newcommand{\pard}{{\rm\partial}}
\renewcommand{\e}{\mathop{\rm e}\nolimits}

\usepackage{iopams} 
\usepackage{cite}  
\usepackage{graphicx}
\usepackage{epstopdf}

\begin{document}

\title[Photon model of light: Revision of applicability limits]{Photon model of light: Revision of applicability limits}

\author{Yuriy~Akimov and Pavlo~Rutkevych\footnote{Present address: Institute of Chemical and Engineering Sciences,
1 Pesek Road, Singapore 627833.}}
\address{Institute of High Performance Computing,
1 Fusionopolis Way, \#16-16 Connexis, Singapore 138632}
\ead{akimov@ihpc.a-star.edu.sg}

\vspace{10pt}
\begin{indented}
\item[]\date{today}
\end{indented}

\begin{abstract}
The photon model of light has been known for decades to be self-inconsistent and controversial theory with numerous intrinsic conflicts. This paper revises the model and explores its applicability for description of classical electromagnetic fields. The revision discloses that the photon model fails for fields in current-containing domains, as well as for near fields in current-free regions. This drastically changes the hierarchy of optics theories and the entire landscape of physics. In particular, quantum optics appears to be not the most advanced theory, as it is commonly thought, but just an improved version of geometrical optics with limited applicability, while quantum electrodynamics turns out to provide a truncated description of electromagnetic interactions.
\end{abstract}

\vspace{2pc}
\section{Introduction}

Although quantum electrodynamics has gained wide acceptance among the scientific community and eventually has led to the formation of all subsequent quantum field theories, it remains one of the most controversial and mathematically inconsistent parts in theoretical physics. 
Regarding it, Paul Dirac, the founder of second quantization, wrote: ``{\it ...most physicists are very satisfied with the situation. 
They say: ``Quantum electrodynamics is a good theory, and we do not have to worry about it any more.'' 
I must say that I am very dissatisfied with the situation, because this so-called ``good theory'' does involve neglecting infinities which appear in its equations, neglecting them in an arbitrary way. 
This is just not sensible mathematics. 
Sensible mathematics involves neglecting a quantity when it turns out to be small -- not neglecting it just because it is infinitely great and you do not want it!}''  \cite{Dirac:1978}
He criticized the treatment of infinite integrals that takes place throughout quantum electrodynamics and that comes right from his theory of second quantization.
Richard Feynman, who has also made a substantial contribution to quantum electrodynamics, in particular by developing renormalization techniques for treatment of the infinite integrals, was never fully satisfied with its mathematical validity either: ``{\it The shell game that we play {\rm [...]} is technically called `renormalization'. But no matter how clever the word, it is still what I would call a dippy process! Having to resort to such hocus-pocus has prevented us from proving that the theory of quantum electrodynamics is mathematically self-consistent. It's surprising that the theory still hasn't been proved self-consistent one way or the other by now; I suspect that renormalization is not mathematically legitimate.}''  \cite{Feynman:1985}
Now, several decades after Dirac and Feynman, the situation in quantum electrodynamics remains unchanged \cite{Cramer:2020}: despite various renormalization techniques proposed, all theories developed lack mathematical rigor and self-consistency \cite {Fraser:2009, Miller:2018} with a number of issues directly coming from second quantization of electromagnetic field \cite{Feynman:2012,Berestetskii:2012}. 
The belief in correctness of theoretical predictions of quantum electrodynamics is ``{\it ultimately based only on their excellent agreement with experiment, not on the internal consistency or logical ordering of the fundamental principles of the theory}'' \cite{Berestetskii:2012}.
This belief pushes the physics community to further develop quantum field theories and related technologies using the mathematically inconsistent model of photons. 
To clarify this situation, we reconsider the main points of the Dirac's second quantization and verify them for validity.

\section{Photon model}

Second quantization of electromagnetic field assumes that there exist elementary indivisible portions of fields, called photons, and that any electromagnetic field can be completely decomposed over such quanta. 
Derivation of photons starts from the Maxwell's equations for free electromagnetic field (in the absence of charges and currents) \cite{Feynman:2012,Berestetskii:2012},
\begin{eqnarray}
	&\displaystyle
	\nabla\times \vect B(t,\vect r) = \varepsilon_0\mu_0\frac{\partial \vect E(t,\vect r)}{\partial t},
\label{Maxwell1-free}\\
	&\displaystyle
	\nabla\times \vect E(t,\vect r) = -\frac{\partial \vect B(t,\vect r)}{\partial t},
\label{Maxwell2-free}
\end{eqnarray}
where $\varepsilon_0=8.85418782 \cdot 10^{-12}$ F/m and $\mu_0=4\pi \cdot 10^{-7}$ H/m are the electric and magnetic constants, $\vect E$ and $\vect B$ are the microscopic electric field and magnetic induction, respectively. 
Solutions of these equations are given as superposition
\begin{eqnarray}
	&\displaystyle
	\vect E(t,\vect r)=\int_0^\infty \vect E_e(t,\vect r;\vect k)\; \d \vect k,
	\label{free fields E total}\\
	&\displaystyle
	\vect B(t,\vect r)=\int_0^\infty \vect B_e(t,\vect r;\vect k)\; \d \vect k
	\label{free fields B total}
\end{eqnarray}
of real-valued plane eigenwaves \cite{Feynman:2012,Berestetskii:2012,Stratton:1941,Landau:1962}
\begin{eqnarray}
	&\displaystyle
	\vect E_e(t,\vect r;\vect k)={\rm Re}\{\vect E_0(\vect k) \exp[\i(\vect k\cdot \vect r-\omega t)]\},
	\label{free fields E}\\
	&\displaystyle
	\vect B_e(t,\vect r;\vect k)={\rm Re}\{(\vect k/\omega)\times\vect E_0(\vect k) \exp[\i(\vect k\cdot \vect r-\omega t)]\}
	\label{free fields B}
\end{eqnarray}
that propagate in all possible directions at angular frequencies $\omega=kc$, where $c=1/\sqrt{\varepsilon_0\mu_0}$ is the speed of light in vacuum. 
The free fields can have varieties of spatial profiles depending on the specified complex amplitude $\vect E_0(\vect k)$, which is an arbitrary function orthogonal to the wavevector, 
\begin{equation}
	\vect E_0(\vect k)\cdot\vect k=0.
\end{equation}

According to the Dirac's second quantization \cite{Feynman:2012,Berestetskii:2012}, there exist quanta of eigenfields (\ref{free fields E}) and (\ref{free fields B}) with two polarizations $\alpha=1,2$ given by an arbitrary pair of orthogonal unit vectors $\vect e_{\alpha}\perp\vect k$. These quanta exhibit the same properties independent of the propagation direction and polarization. 
Following this theory, electromagnetic fields cannot be measured with absolute accuracy due to Heisenberg's uncertainty principle for quantum oscillators.  
However, with increasing number of photons $N_\alpha(\vect k)$ composing the field, the relative uncertainty decreases, such that the classical (absolutely measurable) eigenfields given by Eqs.~(\ref{free fields E}) and (\ref{free fields B}) become valid in the quasi-classical limit of $N_\alpha(\vect k)\gg1$ \cite{Feynman:2012,Berestetskii:2012} with
\begin{equation}
	|\vect E_0(\vect k)\cdot\vect e_\alpha|=\left(\frac{2\hbar c k N_\alpha(\vect k)}{\varepsilon_0}\right)^{1/2}, 
\end{equation}
where $\hbar=1.054571800 \cdot 10^{-34}$ ${\rm J}\cdot{\rm s}$ is the reduced Plank constant. 

\section{Issues of second quantization}

The first issue that arises from the Dirac's theory relates to the total energy of free field, whose value turns out to be absolutely measurable and linear to the number of photons $N_\alpha(\vect k)$ \cite{Feynman:2012,Berestetskii:2012}, 
\begin{eqnarray}
	&\displaystyle
	W=W_0 +\frac{\hbar c}{(2\pi)^3}\sum\limits_{\alpha=1}^2
	\int\limits_0^\infty N_\alpha(\vect k) k \;\d\vect k,
	\label{eq:Wtot}
\end{eqnarray} 
where $W_0$ is the vacuum-state energy in the absence of photons, 
\begin{eqnarray}
	&\displaystyle
	W_0=\frac{\hbar c}{2(2\pi)^3}\sum\limits_{\alpha=1}^2
	\int\limits_0^\infty k \;\d\vect k.
	\label{eq:W0}
\end{eqnarray} 
With the integration performed over all possible wavevectors $\vect k$, the vacuum-state energy $W_0$ appears infinite \cite{Feynman:2012,Berestetskii:2012}. 
As a result, the steady-state energy $W$ of the free field is never finite regardless of the photon numbers $N_\alpha(\vect k)$. 
Eventually, we observe an avalanche of divergent integrals throughout quantum electrodynamics \cite{Feynman:2012,Berestetskii:2012} that forces theoreticians to do mathematically illegitimate manipulations in order to treat the infinities arising in calculations \cite{Dirac:1981}. 

Conventionally, the infinite vacuum-state energy $W_0$ is  treated in two steps. First, it is made finite \cite{Feynman:2012}, by cutting off the wavevector space and limiting the maximum available value of $k$ by a some finite $k_{\rm max}$. 
This procedure is commonly explained by hypothetical failure of Maxwell's equation at small scales: ``{\it we have no evidence that the laws of electrodynamics are obeyed for wavelengths shorter than any which have yet been observed}'' \cite{Feynman:2012}.
Although this argument sounds tenable from the physics point of view, it is just a ``{hocus-pocus}'' from the mathematics perspective.
Before second quantization, the energy of classical free fields (\ref{free fields E total}) and (\ref{free fields B total}),
\begin{equation}
	W=4\pi^3\varepsilon_0\int\limits_0^\infty |\vect E_0(\vect k)|^2\;\d\vect k, 
	\label{W_class}
\end{equation}
had the integral form similar to Eqs.~(\ref{eq:Wtot}) and (\ref{eq:W0}), but did not require any cutting off of the wavevector space for proper dependences of $\vect E_0(\vect k)$. 
However, after second quantization it turns out requiring such truncation.
Regarding the truncated value of $W_0$, there is another issue known in cosmology as the ``vacuum catastrophe'' \cite{Adler:1995}: the discrepancy between theoretical predictions and experimental observations of $W_0$ reaches  120 orders of magnitude, being ``{\it  the worst theoretical prediction in the history of physics}'' \cite{Hobson:2006}. 

The second step in conventional treatment of $W_0$ is ``zeroing'' of its truncated value \cite{Feynman:2012}. 
This manipulation is commonly explained by the ability to choose the zero point of energy arbitrary. 
However, it contradicts the general theory of relativity, where no adjustment of the energy zero is allowed: ``{\it ...unfortunately, it is really not true that the zero point of energy can be assigned completely arbitrary. 
Energy is equivalent to mass, and mass has a gravitational effect. Even light has a gravitational effect.}''  \cite{Feynman:2012}
The only reasonable argument for the forced ``zeroing'' of the trancated $W_0$ is nonobservance of its expected pronounced gravitational effects.
However, being fully empirical, this argument does not add to the mathematical legitimacy of the developed theory.  

Unsurprisingly that such a controversial treatment of the vacuum-state energy is reasonably considered as ``{\it the fact that the present theory is not logically complete and consistent}'' \cite{Berestetskii:2012}, rising the question about applicability limits of the developed model of photons.

\section{Applicability of free-field model}

Although the photon model lacks mathematical rigor and consistency, some of its results are in surprisingly good agreement with experiment.
It seems to give positive results in the area of special relativity all the time: 
to our best knowledge, there are no confirmed failures of quantum electrodynamics in this area reported in the literature. 
This fact gives a great hope to the scientific community that quantum electrodynamics with its photon model is in conflict with gravity theory only and gives meaningful predictions for special theory of relativity. 
But, meaningful to what extent? How universal are these predictions?
Unfortunately, these questions were never investigated thoroughly. 
To address them, we return to the starting point of second quantization procedure.

Recall, the photon model was derived from plane eigenwaves given by Eqs.~(\ref{free fields E}) and (\ref{free fields B}). 
Obviously, it is limited to description of free fields (\ref{free fields E total}) and (\ref{free fields B total}) only. 
Nonetheless, it is considered universal and commonly applied to any arbitrary excited electromagnetic fields \cite{Feynman:2012,Berestetskii:2012}. 
As a result, {\it forced} (excited by currents) electromagnetic fields are treated as superposition of {\it free} field quanta, although forced and free fields obey different differential equations! 
Below, we discuss this point in the quasi-classical limit of $N_\alpha(\vect k)\gg1$. 
Within this limit, superposition of free-field quanta makes sense for the classical forced fields, only if the latter can be completely decomposed over plane eigenwaves similarly to classical free fields (\ref{free fields E total}) and (\ref{free fields B total}). 
In fact, such decomposition is mathematically forbidden. 
We can see that from the equations describing classical forced and free fields. 
Classical forced fields excited by classical or quantum currents with a given density $\vect J(t,\vect r)$ obey the {\it inhomogeneous} Maxwell's equations, 
\begin{eqnarray}
	&\displaystyle
	\nabla\times \vect B(t,\vect r) = \varepsilon_0\mu_0\frac{\partial \vect E(t,\vect r)}{\partial t}+
	\mu_0\vect J(t,\vect r),
\label{Maxwell1}\\
	&\displaystyle
	\nabla\times \vect E(t,\vect r) = -\frac{\partial \vect B(t,\vect r)}{\partial t},
\label{Maxwell2}
\end{eqnarray}
while the free-field model gives only the solutions to the {\it homogeneous} equations with $\vect J(t,\vect r)=0$. 
For the homogeneous Maxwell's equations, any linear combination of its particular solutions (eigenfields) is a solution as well. 
It means that any linear combination of eigenfields (\ref{free fields E}) and (\ref{free fields B}) obeys the Maxwell's equations with $\vect J(t,\vect r)=0$ and can never give the solution for the excited fields in current-containing domains where $\vect J(\vect r,t)\neq0$. 
In other words, it is mathematically impossible to decompose the classical forced fields $\vect E_f(t,\vect r)$ and $\vect B_f(t,\vect r)$ given by Eqs.~(\ref{Maxwell1}) and (\ref{Maxwell2}) over the classical eigenfields $\vect E_e(t,\vect r)$ and $\vect B_e(t,\vect r)$ at  $\vect r\in V_{\vect J\neq 0}$.  

General impossibility of expansion of the forced fields over eigenfields is the crucial conclusion for entire quantum electrodynamics. 
Following it, the linear decomposition of $\vect E_f(t,\vect r)$ and $\vect B_f(t,\vect r)$ over $\vect E_e(t,\vect r)$ and $\vect B_e(t,\vect r)$ makes sense only in current-free domains. 
At the same time, Heisenberg's uncertainty principle for charged particles forbids existence of continuous three-dimensional domains with $\vect J(t,\vect r)=0$. As such, the photon model fails to provide the {\it complete} description for electromagnetic fields in quantum regime of charge dynamics at least in the case of $N_\alpha(\vect k)\gg1$. 

Even in the classical regime of charge dynamics that allows existence of current-free domains, the forced fields $\vect E_f(t,\vect r)$ and $\vect B_f(t,\vect r)$ can be completely decomposed over eigenwaves at $\vect r\in V_{\vect J= 0}$, but only under the proper choice of the fields $\vect E_e(t,\vect r)$ and $\vect B_e(t,\vect r)$. 
Indeed, complete decomposition of the forced fields in a current-free domain $V_{\vect J= 0}$ requires 
a {\it complete set} of eigenfields that must include all solutions of Eqs.~(\ref{Maxwell1-free}) and (\ref{Maxwell2-free}) that are non-divergent inside $V_{\vect J= 0}$ and generally divergent in $V_{\vect J\neq 0}$. 
If we choose an incomplete set of eigenfields, then the decomposition can be performed only within a limited part of the current-free domain. 
For $\vect E_e(t,\vect r)$ and $\vect B_e(t,\vect r)$ chosen in the form of plane waves (\ref{free fields E}) and (\ref{free fields B}), the decomposition for an arbitrary shaped $V_{\vect J= 0}$ is possible only partially, as these eigenfields do not represent a complete set of eigenfunctions. 
It is caused by the {\it zero divergence} of the plane-wave form of $\vect E_e(t,\vect r)$ and $\vect B_e(t,\vect r)$ that does not support existence of the fields' sources at any point of the space (even outside of $V_{\vect J= 0}$).

\section{Plane-eigenwave decomposition}

To illustrate the limitation of plane-eigenwave decomposition of forced fields, we consider a classical Hertzian point dipole oscillating at the frequency $\omega_0$ with the current density
\begin{equation}
	\vect J(t,\vect r)={\rm Re}\left[\vect e_z\delta(\vect r)\e^{-\i\omega_0 t}\right].
\end{equation} 
The fields excited by such a dipole can be written as   
\begin{eqnarray}
	&\displaystyle
	\vect E_f(t,\vect r)={\rm Re}[\nabla\times\nabla\times\vect\Pi_f(t,\vect r)],\label{E_ftr}\\
	&\displaystyle
	\vect B_f(t,\vect r)=\varepsilon_0\mu_0\omega_0\;{\rm Im}[\nabla\times\vect\Pi_f(t,\vect r)],
\end{eqnarray}  
with the use of the Hertz vector [see \ref{app_Hertz} and \ref{app_Dipole} for details]
\begin{equation}
	\vect \Pi_f(t,\vect r)=\vect e_z\frac{\i\e^{\i(k_0 r-\omega_0 t)}}{4\pi \varepsilon_0\omega_0 r},
\end{equation}  
where $k_0=\omega_0/c$. 
The Hertz vector representation is general for any electromagnetic fields \cite{Stratton:1941}. 
Eigenfields (\ref{free fields E}) and (\ref{free fields B}) can be rewritten in this form as well. 
For plane eigenwaves of the angular frequency $\omega_0$, we have 
\begin{eqnarray}
	&\displaystyle
	\vect E_e(t,\vect r)={\rm Re}[\nabla\times\nabla\times\vect\Pi_e(t,\vect r)],\\
	&\displaystyle
	\vect B_e(t,\vect r)=\varepsilon_0\mu_0\omega_0\;{\rm Im}[\nabla\times\vect\Pi_e(t,\vect r)],
\end{eqnarray}  
where the Hertz vector of plane eigenwaves is given by
\begin{equation}
	\vect \Pi_e(t,\vect r)=\vect E_0(|k_0|\vect n)\frac{\e^{\i(|k_0| \vect n\cdot \vect r-\omega_0 t)}}{k_0^{2}}\label{Pi_etr}
\end{equation}  
with $\vect n$ being the unitary vector of the eigenwave propagation [see \ref{app_Eigenwaves} for details].
Writing both fields through Hertz vectors of the same gauge, ${\rm div}~\vect \Pi_f(t,\vect r)={\rm div}~\vect \Pi_e(t,\vect r)= 0$ at $\vect r\in V_{\vect J=0}$, allows us to reduce field decomposition to expansion of $\vect \Pi_f(t,\vect r)$ over $\vect \Pi_e(t,\vect r)$, where $\vect E_0(|k_0|\vect n)$ should be considered as the expansion coefficients. 

\begin{figure}[b]
\begin{center}
\includegraphics[width=0.7\textwidth,clip,trim={1.7cm 0.5cm 3.4cm 1.8cm}]{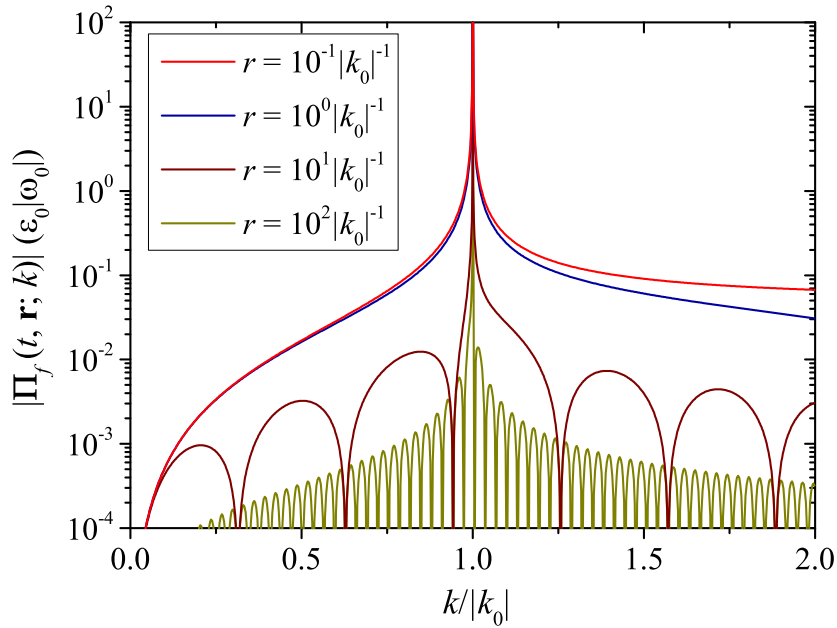}
\end{center}
\caption{Total contributions of the Fourier harmonics with fixed wavenumbers $k$ to $\vect \Pi_{f}(t,\vect r)$ at different distances $r$ from the point dipole. All harmonics with $k\neq |k_0|$ given by non-zero divergence of $\vect \Pi_f(t,\vect r)$ at $\vect r= 0$ are suppressed with growth of $r$. At $r\rightarrow\infty$, only the harmonics with $k=|k_0|$ that correspond to the plain eigenwaves remain in the decomposition. \label{Fig2}}
\end{figure}

To verify completeness of the decomposition, we do Fourier transform of the Hertz vectors to the $\vect k$ space and rewrite them in the integral form over the Fourier harmonics' wavenumbers $k$,
\begin{equation}
	\vect \Pi_{f,e}(t,\vect r)=\int\limits_0^\infty\vect \Pi_{f,e}(t,\vect r;k)\;\d k.
	\label{Pi in k}
\end{equation}  
Here, $\vect\Pi_{f,e}(t,\vect r;k)$ are the integral contributions of all Fourier harmonics with different $\vect k$, but the same absolute value $k$, at a given point $\vect r$:
\begin{equation}
	\vect \Pi_{f,e}(t,\vect r;k)=\frac{1}{(2\pi)^{3/2}}\int\limits_0^{4\pi}\vect \Pi_{f,e}(t,\vect k)k^2\e^{\i\vect k\cdot\vect r}\d\Omega,
\end{equation}  
where $\d\Omega=\sin\theta\;\d\theta\;\d\phi$ is the differential solid angle of the wavevector $\vect k$, and $\vect \Pi_{f,e}(t,\vect k)$ are the Fourier images of the Hertz vectors.
The notation of $\vect \Pi_{f,e}(t,\vect r)$ in the form of (\ref{Pi in k}) clearly demonstrates limitation of the plane eigenwave expansion, as the forced Hertz vector $\vect \Pi_{f}(t,\vect r)$ is broadband in the $k$ space,
\begin{equation}
	\vect \Pi_f(t,\vect r;k)=\vect e_z\frac{\i\e^{-\i\omega_0 t}}{2\pi^2\varepsilon_0\omega_0}\frac{\sin kr}{r}\frac{k}{k^2-k_0^2},
	\label{Pi_f}
\end{equation}  
while the eigen vectors $\vect \Pi_{e}(t,\vect r)$ are completely contributed by the Fourier harmonics with $k=|k_0|$ only,
\begin{equation}
	\vect \Pi_e(t,\vect r;k)=\vect E_0(|k_0|\vect n)\frac{\e^{\i(|k_0|\vect n\cdot \vect r-\omega_0 t)}}{k_0^{2}}\delta(k-|k_0|).
\end{equation}  
The broadbandness of $\vect \Pi_f(t,\vect r;k)$ is caused by discontinuity of the forced fields at $\vect r=0$, where $\vect \Pi_f(t,\vect r)$ experiences non-zero divergence due to the dipole presence,
\begin{equation}
	\left.{\rm div}~\vect \Pi_f(t,\vect r)\right|_{\vect r=0}\neq 0, 
\end{equation}   
which requires inclusion of the Fourier harmonics with $k\neq |k_0|$ for complete plane-wave expansion of the fields. In other words, the decomposition of forced fields over plane eigenwaves (\ref{free fields E}) and (\ref{free fields B}) generally fails, as the latter do not represent a complete set of $\vect E_e(t,\vect r)$ and $\vect B_e(t,\vect r)$ for a non-simply connected current-free space $V_{\vect J= 0}$.

Representation (\ref{Pi in k}) is also useful to investigate contributions of the Fourier harmonics with different wavenumbers $k$ to the forced fields at different points of space $\vect r$. 
According to (\ref{Pi_f}), $\vect \Pi_f(t,\vect r;k)$ exhibits independence of the direction of $\vect r$ and is governed by the distance $r$ only. 
Figure~\ref{Fig2} shows absolute values $|\vect\Pi_f(t,\vect r;k)|$ normalized by $(\varepsilon_0|\omega_0|)^{-1}$ at different distances $r$ from the  dipole position. 
It demonstrates two main points: (i) the forced electromagnetic fields are always broadband in the $k$ space and (ii) all Fourier harmonics with $k\neq |k_0|$ are suppressed, when we move away from the dipole. 
In the limiting case of $r\rightarrow \infty$, 
\begin{equation}
	\lim\limits_{r\rightarrow \infty}\vect \Pi_f(t,\vect r;k)=\vect e_z B\e^{-\i\omega_0 t}\delta(k-|k_0|),
\end{equation}  
where $B$ is an undefined constant given by the $0\cdot\infty$ uncertainty in  Eq.~(\ref{Pi_f}) at $k=|k_0|$. 
Thus, the eigen vectors $\vect \Pi_e(t,\vect r)$ can be used for decomposition of the forced Hertz vector $\vect \Pi_f(t,\vect r)$ at $r\rightarrow \infty$ only. 
For finite distances $r$, it generally fails and, therefore, must be treated as a far-field approximation. 

\section{Limitation of photon model}

Note the fundamental nature of the derived far-field limitation for plane-eigenwave decomposition of dipole radiation. 
As any (classical or quantum) current density $\vect J(t,\vect r)$ 
can be completely decomposed over continuously distributed Hertzian point dipoles, 
\begin{equation}
	\vect J(t,\vect r)=
	\int\limits_{-\infty}^\infty \frac{\d\omega_0}{\sqrt{2\pi}}\int\limits_{V_{\vect J\neq 0}} \d\vect r' \;
	\vect p(\omega_0,\vect r')\delta(\vect r-\vect r')\e^{-\i\omega_0 t},
\end{equation}  
where 
\begin{equation}
	\vect p(\omega_0,\vect r')=\frac{1}{\sqrt{2\pi}}\int\limits_{-\infty}^\infty \vect J(t,\vect r')\e^{\i\omega_0 t}\d t
\end{equation}
is the local dipole strength at the point $\vect r=\vect r'$ with $\omega=\omega_0$, the electric fields $\vect E_f(t,\vect r)$ excited by $\vect J(t,\vect r)$ 
can be written through the integral dipole Hertz vector \cite{Stratton:1941},
\begin{equation}
	\vect \Pi_f(t,\vect r)=\i
	\int\limits_{-\infty}^\infty \frac{\d\omega_0}{\sqrt{2\pi}}\int\limits_{V_{\vect J\neq 0}} \d\vect r' \;
	\vect p(\omega_0,\vect r')\frac{\e^{\i(k_0 |\vect r-\vect r'|-\omega_0 t)}}{4\pi\varepsilon_0\omega_0 |\vect r-\vect r'|}.
\end{equation}   
As such, the far-field limitation of plane-eigenwave decomposition derived for a single dipole naturally appears for any 
distribution of $\vect J(t,\vect r)$.  

Thus, the description of electromagnetic interactions through exchange of photons used in quantum electrodynamics represents the truncated form of interaction, where only the Fourier harmonics with $k=|k_0|$ corresponding to far fields are accounted, but other harmonics related to near fields are ignored. 
Such a description is valid, only if the distance between the interacting currents is much larger than the wavelength, $2\pi/|k_0|$. 
It restricts the results given by quantum electrodynamics to the limit $|k_0|\rightarrow \infty$, and, hence, requires their revision and rethinking in this regard. 

\section{Hierarchy of optics theories}

Also, reconsideration is required for the hierarchy of optics theories.
Remind, the modern landscape consists of three main theories: (i) geometrical optics, (ii) wave optics, and (iii) quantum optics. 
In the existing hierarchy, geometrical optics is considered as the simplest approach, where light is described in terms of rays, and its wave nature is ignored; wave optics is treated as intermediate theory, where light is given by electromagnetic fields without assumption of their quantization; and  quantum optics is thought to be the most advanced and sophisticated theory, where light is described with an ensemble of photons that exhibit both wave and particle properties.
Now, we can rearrange this hierarchy based on the above revision of the photon model. 

Geometrical optics remains the well-established reduction of wave optics in the limit $|k_0|\rightarrow \infty$, with focus on the far-field zone \cite{Landau:1962}.
It naturally loses all information about wave nature of light and is unable to describe diffraction, interference, and radiation in the near-field zone close to the sources.
Rays in geometrical optics are uncategorized; polarization or color can be added only phenomenologically.

Regarding quantum optics, we need to address both the particle and wave properties of light. 
On the particle level, quantum optics describes light as an ensemble of delocalized noninteracting quasiparticles of fixed energy and momentum. 
On the wave level, it considers light as plane eigenwaves propagating in the far-field zone. 
Thus, the applicability limit of quantum optics for the wave description of light appears identical to that of geometrical optics. 
The only difference is that quantum optics makes a try to account for the wave nature of light by assigning frequency, wavenumber, and polarization  to photons, in addition to their particle properties given by geometrical optics \cite{Landau:1962}. 
As such, it enables the description of diffraction and interference in the far-field zone, but generally fails in the near-field zone. 

From the perspective of classical (wave) description, the revised hierarchy of optics theories looks as follows: (i) geometrical optics as the simplest level of description, (ii) quantum optics on the intermediate level, and (iii) wave optics on the highest level as the most sophisticated theory. 
As for description of non-classical (quantum) effects, quantum optics remains pretending to be advantageous (as far as possible for a mathematically inconsistent theory) but limited to far-fields, while wave optics also pretends to have similar capability but without the restriction to far-fields, as follows from various revisions of classical field theory \cite{Lamb:1995,Cercignani:1998,Camparo:1999,Rashkovskiy:2015-1,Rashkovskiy:2016}.

\section{Conclusion}

To conclude, the advantage of quantum electrodynamics over classical description is a wide-spread belief that still remains unproven. 
There is no evidence that quantum description of electromagnetic fields completely covers the classical one for increasing number of photons. 
In fact, it fails for full-wave description of classical fields, as has been demonstrated in this paper.  
As such, quantum electrodynamics cannot be considered as generalization of classical electrodynamics to an upper level of description. 
This conclusion drastically changes the landscape of modern physics and requires further verification of all quantum field theories in view of their limited applicability. 
Particularly, most of quantum calculations for absorption and emission of light go beyond the far-field applicability limit of the photon model and, thus, must be revised.  
In addition, a new revision of particle physics with its Standard Model is required, as photons appear to be not fundamental quasiparticles.

\appendix

\section{Hertz vector formalism}\label{app_Hertz}

The Hertz vector formalism is an old-fashioned approach that allows to write the excited fields in a compact and elegant form. 
In the literature, there exist slightly different definitions of the Hertz vector (see, for example, Refs.~\cite{Stratton:1941,Lucosz:1977}). To avoid confusion, 
we briefly derive the results presented in Eqs.~(\ref{E_ftr})--(\ref{Pi_etr}) of the paper.
For brevity we omit the arguments $(t, \vect{r})$ in all field functions.

Conventionally \cite{Stratton:1941,Landau:1962}, solutions of the Maxwell's equations (\ref{Maxwell1}) and (\ref{Maxwell2}) are sought with the scalar and vector potentials $\phi$ and $\vect{A}$,
\begin{equation}\label{sup:eq:A}
	\vect B = \nabla \times \vect A ,\qquad
	\displaystyle\vect E = - \frac{\pard \vect A}{\pard t} - \nabla \phi.
\end{equation}
As the potentials are not uniquely defined, there exist different gauges that enable finding the potentials from different differential equations. For time-varying fields, the most popular one is the Lorentz gauge imposing the following relation between the vector and scalar potentials  \cite{Stratton:1941,Landau:1962}, 
\begin{equation}
\label{sup:eq:Lorentz}
\nabla\cdot \vect{A} + \varepsilon_0 \mu_0 \frac{\pard \phi}{\pard t} = 0.
\end{equation}
In this gauge, the Maxwell's equations for electric field and magnetic induction split into two similar wave equations for $\vect A$ and $\phi$,
\begin{eqnarray}
\label{sup:eq:Maxwell:2-1}
&\displaystyle\nabla^2 \vect{A} - \varepsilon_0 \mu_0 \frac{\pard^2 \vect{A}}{\pard t^2} 
= -\mu_0 \vect{J},
\\
\label{sup:eq:Maxwell:2-2}
&\displaystyle\nabla^2 \phi - \varepsilon_0 \mu_0 \frac{\pard^2 \phi}{\pard t^2} 
= -\frac{1}{\varepsilon_0} \rho.
\end{eqnarray}
If we let the vector and scalar potentials be expressed through the Hertz vector $\vect \Pi$ as follows,
\begin{equation}
\label{sup:eq:HertzA}
\vect{A}= \varepsilon_0 \mu_0 \frac{\pard \vect{\Pi}}{\pard t},\qquad
\phi = - \nabla \vect{\Pi},
\end{equation}
then Lorentz gauge condition (\ref{sup:eq:Lorentz}) is satisfied for any $\vect \Pi$.
Substituting these definitions into (\ref{sup:eq:Maxwell:2-1}) and (\ref{sup:eq:Maxwell:2-2}), we obtain
the single equation for the Hertz vector:
\begin{equation}
\label{sup:eq:Maxwell:Pi}
\frac{\pard}{\pard t}\left(\nabla^2 \vect{\Pi} - \varepsilon_0 \mu_0 \frac{\pard^2 \vect{\Pi}}{\pard t^2}\right) 
= -\frac{\vect{J}}{\varepsilon_0}.
\end{equation}
This equation defines the Hertz vector together with 
the magnetic induction and electric field,
\begin{equation}
\label{sup:eq:HertzB_tot}
\vect{B}= \varepsilon_0 \mu_0 \nabla \times \frac{\pard\vect{\Pi}}{\pard t},\qquad
\vect E = \nabla^2 \vect \Pi- \varepsilon_0 \mu_0 \frac{\pard^2 \vect{\Pi}}{\pard t^2}.
\end{equation}

The Hertz vector is not uniquely defined, as can be seen from Eq.~(\ref{sup:eq:Maxwell:Pi}). Indeed, if we replace $\vect \Pi$ with $\vect \Pi+\vect \Psi$ there, it remains valid for any $\vect \Psi$ satisfying
\begin{equation}
\nabla^2 \vect{\Psi} - \varepsilon_0 \mu_0 \frac{\pard^2 \vect{\Psi}}{\pard t^2}={\rm const}.
\end{equation}
It is convenient to define the Hertz vector in the gauge, when ${\rm div~}\vect\Pi=0$ at $\vect r\in V_{\vect J=0}$. In this gauge, the Hertz vector in a current-free domain can be obtained from the equation
\begin{equation}
\label{sup:eq:Maxwell:Pi-free}
\nabla^2 \vect{\Pi} - \varepsilon_0 \mu_0 \frac{\pard^2 \vect{\Pi}}{\pard t^2}=0,
\end{equation} 
with the the magnetic induction and electric field given by
\begin{equation}
\label{sup:eq:HertzB}
\vect{B}= \varepsilon_0 \mu_0 \nabla \times \frac{\pard\vect{\Pi}}{\pard t},
\qquad
\vect{E}= \nabla\times\nabla\times\vect{\Pi}.
\end{equation}

\section{Description of a Hertzian point dipole}\label{app_Dipole}

To get the fields generated by a point dipole with the current density  
\begin{equation}
	\vect J=\vect e_z\delta(\vect r)\e^{-\i\omega_0 t},
\end{equation} 
we can solve vector Hertz equation (\ref{sup:eq:Maxwell:Pi})
\begin{equation}
\frac{\pard}{\pard t}\left(\nabla^2 \vect{\Pi} - \varepsilon_0 \mu_0 \frac{\pard^2 \vect{\Pi}}{\pard t^2}\right) 
= -\frac{1}{\varepsilon_0}\vect e_z\delta(\vect r)\e^{-\i\omega_0 t}.
\end{equation}
In the gauge ${\rm div~}\vect\Pi=0$ at $\vect r\in V_{\vect J=0}$, it can be rewritten as follows
\begin{equation}
\nabla^2 \vect{\Pi} - \varepsilon_0 \mu_0 \frac{\pard^2 \vect{\Pi}}{\pard t^2} 
= -\frac{\i}{\varepsilon_0\omega_0}\vect e_z\delta(\vect r)\e^{-\i\omega_0 t}.
\end{equation}
Solution of this equation is given by 
\begin{equation}
	\vect \Pi=\vect e_z\frac{\i\e^{\i(k_0 r-\omega_0 t)}}{4\pi \varepsilon_0\omega_0 r}.
\end{equation}  
This is the Hertz vector of a point dipole that can be used together with Eqs.~(\ref{sup:eq:HertzB}) for calculation of the generated fields. 

\section{Description of plane eigenwaves}\label{app_Eigenwaves}

To describe plane eigenwaves with the Hertz vector in the gauge ${\rm div~}\vect\Pi=0$, we need to solve
\begin{equation}
\nabla^2 \vect{\Pi} - \varepsilon_0 \mu_0 \frac{\pard^2 \vect{\Pi}}{\pard t^2} = 0.
\end{equation}
This equation gives us 
\begin{equation}
\vect \Pi=\vect\Pi_0(\vect n)\e^{\i(|k_0| \vect n\cdot \vect r-\omega_0 t)},
\end{equation}
for the fields oscillating at frequency $\omega_0$, where $\vect\Pi_0(\vect n)$ is an arbitrary complex amplitude of the eigenwaves propagating in the direction specified by unitary vector $\vect n$. The gauge condition requires $\vect\Pi_0(\vect n)$ to be perpendicular to the propagation direction $\vect n$, 
\begin{equation}
\vect\Pi_0(\vect n)\cdot\vect n=0.
\end{equation}
In Eq.~(\ref{Pi_etr}), $\vect\Pi_0(\vect n)$ was chosen to be $\vect E_0(|k_0|\vect n) k_0^{-2}$. The eigenfields corresponding to this Hertz vector can be obtained with the use of Eqs.~(\ref{sup:eq:HertzB}).

\end{document}